\def\rfr#1{Eq. (\ref{#1})}

\def\bm#1{{\mbox{\boldmath$#1$\unboldmath}}}

\def\bar{\begin{eqnarray}}
\def\ear{\end{eqnarray}}

\def\eqi{\begin{equation}}
\def\eqf{\end{equation}}
\def\eqia{\begin{eqnarray}}
\def\eqfa{\end{eqnarray}}
\def\rp#1#2{{#1\over#2}}

\def\lb#1{\label{#1}}

\def\oc2{$\mathcal{O}(c^{-2})$}

\documentclass[11pt]{article}
\usepackage{amsmath,amsthm,amscd,amssymb}
\usepackage{latexsym}
\usepackage{graphicx,epsfig}

\begin{document}

\noindent{\bf \LARGE{A critical analysis of a recent test of the
Lense-Thirring effect with the LAGEOS satellites}}
\\
\\
\\
{L. Iorio,  }\\
{\it Viale Unit$\grave{a}$ di Italia 68, 70125\\Bari, Italy
\\tel./fax 0039 080 5443144
\\e-mail: lorenzo.iorio@libero.it}

\begin{abstract}
In this paper we quantitatively discuss the impact of the current uncertainties
in the even zonal harmonic coefficients $J_{\ell}$ of the
Newtonian part of the terrestrial gravitational potential on the
measurement of the general relativistic Lense-Thirring effect
using a suitable linear combination of the nodes $\Omega$ of the
laser-ranged LAGEOS and LAGEOS II satellites. The 1-sigma
systematic error due to the mismodelling in the $J_{\ell}$
coefficients ranges from $\sim 4 \%$ for the EIGEN-GRACE02S
gravity field model to $\sim 9\%$ for the GGM02S
model. Another important source of systematic error of
gravitational origin is represented by the secular variations
$\dot J_{\ell}$ of the even zonal harmonics. While the
relativistic and $J_{\ell}$ signals are linear in time, the shift
due to $\dot J_{\ell}$ is quadratic. We quantitatively assess
their impact on the measurement of the Lense-Thirring effect
with numerical simulations
%To
%this aim, we simulate the time series of the combined LAGEOS and
%LAGEOS II nodes by including the Lense-Thirring signal as
%predicted by the general theory of relativity and the other
%classical dynamical features which affect the orbits of the LAGEOS
%satellites (static and time-dependent parts of the even zonal
%harmonics, tides, non-gravitational perturbations). In order to
%evaluate the systematic error due to $\dot J_{\ell}$ on the
%measurement of the gravitomagnetic shift, we fit our model with a
%straight line only and with a quadratic polynomial and compare the
%slopes of the so-obtained linear trends. We perform 5000 runs, in
%which we randomly vary the noise, the amplitudes and the initial
%phases of the constituents of the model in order to simulate
%different initial conditions and mismodelling levels within the
%currently accepted ranges.
obtaining a  $10-20\%$ 1-sigma total error over 11 years for EIGEN-GRACE02S. Ciufolini and Pavlis,
in a test performed in 2004, claim
a total error of $5\%$ at 1-sigma level.

\end{abstract}

Keywords: Lense-Thirring effect; LAGEOS satellites; GRACE Earth
gravity field models; even zonal harmonics and their secular
variations

%\newpage
%\tableofcontents
%\newpage
\section{Introduction}
\subsection{The Lense-Thirring effect}
The post-Newtonian Lense-Thirring effect (Lense and Thirring 1918,
Soffel 1989, L\"{a}mmezahl and Neugebauer 2001) is one of the few
predictions of the Einsteinian General Relativity Theory (GRT) for
which a direct and undisputable test is not yet available.

According to Einstein, the action of the gravitational potential
$U$ of a given distribution of mass-energy is described by the
metric coefficients $g_{\mu\nu},\ \mu,\nu=0,1,2,3$ of the
space-time metric tensor. They are determined, in principle, by
solving the fully non-linear field equations of GRT for such a
mass-energy content. These equations can be linearized in the
weak-field ($U/c^2 <<1$, where $c$ is the speed of light in
vacuum) and slow-motion ($v/c<<1$) approximation (Mashhoon 2001;
Ruggiero and Tartaglia 2002), valid throughout the Solar System,
and look like the equations of the linear Maxwellian
electromagnetism. Among other things, a noncentral, Lorentz-like
force \eqi\bm F_{\rm LT}=-2m\left(\rp{\bm v}{c}\right)\times \bm
B_g\lb{fgm}\eqf acts on a moving test particle of mass $m$. It  is
induced by the post-Newtonian component $\bm B_g$ of the
gravitational field in which the particle moves with velocity \bm
v. $\bm B_g$ is related to the mass currents of the source of the
gravitational field and comes from the off-diagonal components
$g_{0i}, i=1,2,3$ of the metric tensor. Thanks to such an analogy,
the ensemble of the gravitational effects induced by mass
displacements is also named gravitomagnetism. Far from a central
rotating body of proper angular momentum \bm L the gravitomagnetic
field is\eqi\bm B_g=\rp{G[3\bm r(\bm r\cdot \bm L )-r^2 \bm
L]}{cr^5}.\lb{gmfield}\eqf

One of the consequences of \rfr{fgm} and \rfr{gmfield} is a
gravitational spin--orbit coupling. Indeed, if we consider the
orbital motion of a particle in the gravitational field of a
central spinning mass,  it turns out that
%the
%spin \bm s of the spinning orbiting particle undergoes a tiny
%precessional motion \citep{Schiff 1960}. The most famous
%experiment devoted to the measurement, among other things, of such
%effect in the gravitational field of Earth is the GP--B mission
%\citep{Everitt et al 2001} which has been launched on April 2004.
%
%If the whole orbit of a test particle in its geodesic motion
%around $M$ is considered as a sort of giant gyroscope,
the orbital angular momentum \bm \ell\ of the particle undergoes
the Lense--Thirring precession, so that the longitude of the
ascending node $\Omega$ and the argument of pericentre $\omega$ of
the orbit of the test particle are affected by tiny secular rates
$\dot\Omega_{\rm LT}$, $\dot\omega_{\rm LT}$ (Lense and Thirring
1918, Cugusi and Proverbio 1978, Soffel 1989, Ashby and Allison 1993, Iorio 2001a)
\begin{equation}
\dot\Omega_{\rm LT} =\frac{2GL}{c^2 a^3(1-e^2)^{\frac{3}{2}}},\
\dot\omega_{\rm LT} =-\frac{6GL\cos i}{c^2
a^3(1-e^2)^{\frac{3}{2}}},\lb{LET}
\end{equation} where $a,\ e$ and $i$ are the semimajor axis, the eccentricity and the inclination, respectively, of
the orbit and $G$ is the Newtonian gravitational constant. Note
that in their original paper Lense and Thirring (1918) used the
longitude of the pericentre $\varpi$.

In April 2004 the GP-B mission (Everitt et al. 2001) has been
launched. One of its goals is the measurement of another effect
induced by the terrestrial angular momentum, i.e. the
gravitomagnetic precession of the spins (Schiff 1960) of four
superconducting gyroscopes carried onboard with an expected
accuracy of $1\%$ or better.

However, according to Nordtvedt (2003), the multi-decade analysis
of the Moon's orbit by means of the Lunar Laser Ranging (LLR)
technique yields a comprehensive test of the various parts of
order $\mathcal{O}(c^{-2})$ of the post-Newtonian equation of
motion. The existence of the Lense-Thirring signature as predicted
by GRT would, then, be indirectly inferred from the high accuracy
of the lunar orbital reconstruction. Also the radial motion of the
LAGEOS satellite would yield another indirect confirmation of the
existence of the Lense-Thirring effect (Nordtvedt 1988).

\subsection{Aim of the paper}
In the case of the Earth, the precessions of \rfr{LET} are very
tiny: for the laser-ranged LAGEOS satellites they amount to a few
tens of milliarcseconds per year (mas yr$^{-1}$ in the following).
Extracting such a minute  signal from the background of the much
larger competing classical effects of gravitational (even zonal
harmonics and their secular variations, tides) and
non-gravitational (direct solar radiation pressure, Earth albedo,
direct Earth infrared radiation, thermal forces like the solar
Yarkovsky-Schach effect and the terrestrial Yarkovsky-Rubincam
effect) origin is a very challenging task that requires detailed
knowledge also from many branches like geodesy, geodynamics and
celestial mechanics other than relativistic physics. In turn, the
extremely high accuracy required for such relativistic tests in
the terrestrial space environment can help in further increasing
our knowledge of many aspects traditionally treated by geophysical
and space sciences. In this particular case, as we will see, it is
especially true for  the even zonal harmonic coefficients $
J_{\ell}$ of the terrestrial gravitational potential and their
secular variations $\dot J_{\ell}$.

Recently, a test of the  Lense-Thirring effect on the orbit of a
test particle has been performed by Ciufolini and Pavlis (2004).
They analyzed the data of the laser-ranged LAGEOS and LAGEOS II
satellites in the gravitational field of the Earth by using an
observable proposed in Iorio and Morea (2004). The data analysis
spans 11 years; the 2nd generation GRACE-only EIGEN-GRACE02S Earth
gravity model (Reigber et al. 2005a) released by the
GeoForschungsZentrum (GFZ), Potsdam was adopted. The total
accuracy claimed by Ciufolini and Pavlis is 5-10$\%$ at 1-3 sigma,
respectively, but such estimate is controversial (Iorio 2005a) for
various reasons. One of the major critical points is the impact of
the systematic error induced by the secular variations $\dot
J_{\ell}$ of the even zonal harmonics. Another critical remark is
that only one Earth gravity model was used by Ciufolini and Pavlis
in their analysis.

In this paper, we quantitatively investigate these issues. It will
be shown that the error budget analysis performed by Ciufolini and
Pavlis is optimistic. Our evaluation of the total error in the
Lense-Thirring measurement ranges from 19$\%$ to 24$\%$ at 1-sigma
level.
\section{The LAGEOS-LAGEOS II $J_2$-free combination}
The adopted observable is the following combination of the
residuals $\delta\dot\Omega_{\rm obs}$ of the rates of the nodes
of LAGEOS and LAGEOS II (Iorio and Morea 2004)
\eqi\delta\dot\Omega_{\rm obs}^{\rm
LAGEOS}+c_1\delta\dot\Omega_{\rm obs}^{\rm LAGEOS\ II}\sim
\mu_{\rm LT }S_{\rm LT},\lb{iorform}\eqf where $c_1\sim 0.546$,
$S_{\rm LT}=48.1$ mas yr$^{-1}$ is the slope of the secular trend
according to GRT, and $\mu_{\rm LT}$ is equal to 1 in Einsteinian
theory and 0 in Newtonian mechanics. For the explicit expression
of $c_1$ see below \rfr{coff}. The combination of \rfr{iorform}
has been built up in order to cancel out the first even zonal
harmonic $J_2$, along with its time-varying part which also
includes its secular variations $\dot J_2$. The impact of the
non-gravitational perturbations on the nodes of the LAGEOS
satellites  is of the order of $\sim 1\%$ of the Lense-Thirring
effect (Lucchesi 2001; 2002; 2003; 2004; Lucchesi et al. 2004).

The  idea of using only the nodes of the LAGEOS satellites to
disentangle the Lense-Thirring effect from $J_2$ was proposed in
Ries et al. (2003) in the context of the expected improvements in
our knowledge of the Earth's gravity field from the dedicated
GRACE mission. The linear combination approach was adopted for the
first time by Ciufolini (1996) for his early, less precise tests
with the nodes of the LAGEOS satellites and the perigee of LAGEOS
II (Ciufolini et al. 1998).

Let us write the observed residuals of the rate of the node of a
satellite $\delta\dot\Omega$ in terms of the Lense-Thirring
precession and of the classical precession induced by the Earth's
quadrupole mass moment \eqi\delta\dot\Omega_{\rm
obs}=\dot\Omega_{\rm LT}\mu_{\rm LT}+\dot\Omega_{.2}J_2+[{\rm
higher\ degree\ terms}],\lb{rate}\eqf where $\dot\Omega_{\rm LT}$
is as in \rfr{LET} and
\eqi\dot\Omega_{.2}\equiv\rp{\partial\dot\Omega_{\rm geopotential
}}{\partial J_2}=-\rp{3}{2}n\left(\rp{R}{a}\right)^2\rp{\cos i
}{(1-e^2)^2},\eqf in which $R$ and $n$ are the Earth's mean
equatorial radius and the Keplerian mean motion $n=\sqrt{GM/a^3}$,
respectively. The coefficients $\dot\Omega_{.\ell}$ have been
explicitly worked out in Iorio (2003) up to degree $\ell=20$;
their numerical values for LAGEOS and LAGEOS II, whose orbital
parameters are listed in Table \ref{para}, can be found in Table
\ref{echecaz}. If we write \rfr{rate} for both LAGEOS and LAGEOS
II and solve for $\mu_{\rm LT}$ the so-obtained linear system of
two equations in the two unknowns $\mu_{\rm LT}$ and $J_2$  it is
possible to obtain \rfr{iorform} with \eqi
c_1\equiv-\rp{\dot\Omega^{\rm LAGEOS}_{.2}}{\dot\Omega^{\rm
LAGEOS\ II }_{.2}}=-\rp{\cos i_{\rm LAGEOS}}{\cos i_{\rm LAGEOS\
II}}\left(\rp{1-e^2_{\rm LAGEOS\ II}}{1-e^2_{\rm
LAGEOS}}\right)^2\left(\rp{a_{\rm LAGEOS\ II}}{a_{\rm LAGEOS
}}\right)^{7/2}.\lb{coff}\eqf Note that $c_1$ is independent of
the time span and of the even zonal harmonics: it is only fixed by
the orbital parameters $a,e,i$ of LAGEOS and LAGEOS II.
\section{The impact of the even zonal harmonics}
The combination of \rfr{iorform} is, by construction, independent
of the perturbing effects of degree $\ell=2$ and order $m=0$. In
the case of $J_2$ it can be checked by using the figures in Table
\ref{echecaz} in order to calculate \eqi\dot\Omega_{.2}^{\rm
LAGEOS}+c_1\dot\Omega_{.2}^{\rm LAGEOS \ II};\eqf the result is
zero. Note that the same also holds for the time-dependent
perturbations like the $\ell=2,\ m=0$ constituent of the 18.6-year
tide. Indeed, by calculating the left-hand side of \rfr{iorform}
with the values of the perturbing amplitudes of Table I and Table
II in Iorio (2001b) it is possible to explicitly check this fact.

\subsection{The error due to the static part of the geopotential}
\rfr{iorform} is affected by all the remaining even zonal
harmonics of degree higher than two $J_4, J_6, J_8,...$, along
with their secular variations $\dot J_4, \dot J_6,...$.

The secular rate  induced by $J_4, J_6,...$ can be calculated as
\eqi\sum_{\ell\geq 4}\left(\dot\Omega_{.\ell}^{\rm
LAGEOS}+c_1\dot\Omega_{.\ell}^{\rm LAGEOS\ II}\right)
J_{\ell}\lb{rappos}\eqf by using Table \ref{echecaz} for
$\dot\Omega_{.\ell},\ell=4,6,...$ and the values for $J_{\ell\geq
4 }$ of the chosen Earth gravity field model. Such aliasing shift
is quite larger than the Lense-Thirring rate. Indeed, according
to, e.g., the EIGEN-CG01C (Reigber et al. 2005b) and EIGEN-CG03C
(F\"{o}rste et al. 2005) Earth gravity models, which combine data
from CHAMP, GRACE and terrestrial measurements, the nominal rate
induced by the static part of the even zonals on the combination
of \rfr{iorform} is $\sim 10^5$ mas yr$^{-1}$. It turns out that
for the LAGEOS satellites the precessions due to the zonals with
$\ell\geq 16$ can safely be neglected: indeed, their effect on the
combination of \rfr{iorform} amounts to 0.05 mas yr$^{-1}$; the
sensitivity is $\sim 1$ mas yr$^{-1}$. Thus, it is necessary to
model the action of the low-degree even zonal part of the
classical terrestrial gravitational field very accurately. To be
more precise, in order to reduce the classical aliasing rates to a
level $\sim 1\%$ of the Lense-Thirring effect it would be
necessary to know $J_4$ and $J_6$ with an uncertainty of better
than $2\times 10^{-12}$ and $4\times 10^{-12}$, respectively. The
present-day errors in $J_4$ and $J_6$ are, instead, $8.4\times
10^{-12}$ and $6.6\times 10^{-12}$, according to the latest model
EIGEN-CG03C.

Unfortunately, it turns out that the present-day knowledge of the
even zonal harmonics, represented by the latest models based on
CHAMP and, especially, GRACE, is not yet good enough to allow for
a reliable, model-independent $1\%$ test of the Lense-Thirring
effect with the combination of \rfr{iorform}. Indeed, at present
GRACE seems to experience some difficulties in getting notable
improvements in measuring the even zonal harmonics of low-degree
(Wahr et al. 2004), contrary to the medium-high degree even zonals
which, instead, do not pose problems to \rfr{iorform}. As
summarized in Table \ref{geopo}, the systematic percent bias
$\delta\mu_{\rm LT}^{\rm geopotential}$ induced by the mismodelled
part of the geopotential on the gravitomagnetic effect ranges from
$\sim 4\%$ for EIGEN-CG03C and EIGEN-GRACE02S to $\sim 9\%$ for
GGM02S (Tapley et al. 2005). These figures are 1-sigma upper
bounds obtained by linearly adding the absolute values of the
individual mismodelled precessions, according to \rfr{rappos} with
$\sigma_{J_{\ell\geq 4}}$ instead of $J_{\ell\geq 4}$, and then
compared to the Lense-Thirring rate $S_{\rm LT}$. Since the
signature of such source of bias is the same as that of the
Lense-Thirring effect, it is not possible to fit and remove it
from the time series without also affecting the relativistic
signal of interest. It is only possible to evaluate as accurately
as possible its aliasing impact on the Lense-Thirring trend.

\subsection{The error due to the secular variations of the even zonal harmonics}
Another source of non-negligible systematic error is represented
by the secular changes $\dot J_{\ell\geq 4}$ of the even zonal
harmonics of low-degree (Cheng et al. 1997; Bianco et al. 1998;
Cox and Chao 2002; Dickey et al. 2002; Cox et al. 2003; Cheng and
Tapley, 2004). They are, at present, rather poorly known, as can
be inferred from, e.g., Table 1 by Cox et al. (2003). Such
temporal variations in the even zonals are mainly related to the
Earth's lower mantle viscosity features (Ivins et al. 1993). This
topic has recently received attention mainly due to the observed
inversion of the rate of change of the Earth's quadrupole mass
moment coefficient $J_2$ which, since 1998, began increasing (Cox
and Chao 2002).  It is not yet clear if such an effect is a
long-term feature or is short-term in nature; however, $\dot J_2$
is not relevant in the present context because \rfr{iorform} is
not sensitive to it. However, the possibility of interannual
variations of $J_4$ and $J_6$, depending also on the time span of
the particular solution adopted, cannot be ruled out; they would
affect, in principle, \rfr{iorform}. In Table \ref{jdots} we quote
the weighted means of the best estimates of Table 1 by Cox et al.
(2003) for $\dot J_{\ell\geq 4}$. The uncertainties $\sigma_{\dot
J_{\ell\geq 4}}$ reported in Table \ref{jdots} are the variances
$1/\sigma^2=\sum_i(1/\sigma^2_i)$ of the distributions of the
formal errors $\sigma_i$ of Table 1 by Cox et al. (2003). Note
that the so obtained $\sigma_{\dot J_{\ell\geq 4}}$ are smaller
than such formal errors.

In the following we will perform quantitative evaluations of the
impact of $\dot J_{\ell\geq 4}$ on \rfr{iorform}. As a first,
preliminary approximation, we will use the results of Table
\ref{jdots} which smooth out the possible interannual variations
in the even zonals of interest.

The effect of $\dot J_{\ell}$ integrated over the time grows
quadratically. A first, a priori, quantitative evaluation of the
aliasing impact of $\dot J_4, \dot J_6$ on the recovery of the
Lense-Thirring effect can be performed by calculating the
following quantity over a chosen observational time span $T_{\rm
obs}$ \eqi\rp{\sum_{\ell=4}^6\left(\dot\Omega_{.\ell}^{\rm
LAGEOS}+c_1\dot\Omega_{.\ell}^{\rm LAGEOS\
II}\right)\left(\rp{\dot J_{\ell}}{2}\right)T_{\rm obs}^2}{S_{\rm
LT }T_{\rm obs}}.\lb{rappo}\eqf For $T_{\rm obs}=11$ years, it
turns out that, according to Table \ref{jdots}, their impact
amounts to $\sim 12\%$. Note that since $\dot J_2$ does not affect
the combination of \rfr{iorform} and the errors in $\dot J_4$ and
$\dot J_6$ are of the same order of magnitude of the nominal
values, the situation does not change if we calculate \rfr{rappo}
with the $\sigma_{\dot J_{\ell}}$ instead of their best estimates.
\section{Numerical simulations}
In this Section we investigate the impact of $\sigma_{J_{\ell}}$ and $\sigma_{\dot J_{\ell}}$
by means of numerical analyses.
\subsection{The simulated data}
Following Pavlis and Iorio (2002), we simulate with MATLAB a
time-series of the combined node residuals of the LAGEOS satellites data (called Input Model, IM) for the combination
of \rfr{iorform}  by including in it
\begin{itemize}
\item
LT$\equiv S_{\rm LT}t$. Lense-Thirring trend as predicted by GRT
according to \rfr{iorform}
\item
ZONDOT$\equiv \sum_{\ell=4}^6
r_{\ell}\left(\dot\Omega_{.\ell}^{\rm LAGEOS}+c_1
\dot\Omega_{.\ell}^{\rm LAGEOS\ II} \right)\left(\rp{ \sigma_{\dot
J_{\ell}} }{2}\right)t^2$. Quadratic term due to the $\dot
J_{\ell}$ according to Table \ref{jdots}. The numbers $r_{\ell}$
are randomly generated from a normal distribution with mean zero
and unit standard deviation.
\item
ZONALS$\equiv pS_{\rm LT}\rp{x}{100}t$. Linear trend with a slope
of $x\%$ of the Lense-Thirring according to Table \ref{geopo}. The
number $p$ is randomly generated as $r_{\ell}$.
\item
TIDE$\equiv\sum
a_c\sigma_{A_c}\cos\left[\left(\rp{2\pi}{P}\right)t+f_c\right]$
%+
%\sum\{a_s\}\sigma_{A_s}\sin\left[\left(\rp{2\pi}{P}\right)t+\{f_s\}\right]$.
Set of various tidal perturbations of known periods $P$. For the
impact of such kind of perturbations on the orbits of the LAGEOS
satellites see Iorio (2001b). The numbers $a_c, f_c$ are randomly
generated as $p$ and $r_{\ell}$. The tidal constituents included
are $K_1$, $P_1$, $S_2$, $165.565$ and the 18.6-year tide
\item
NOISE. We also included the effect of various other sources of
non-systematic errors by using a generator of random numbers for
all the points of the simulated time series. The resulting noise
was a gaussian one and the amplitude could be varied.
%as $N\times randn(t),\
%0<t<T_{\rm obs}$. $N$ is an amplitude which can be arbitrarily
%varied.
\end{itemize}
In a nutshell \eqi {\rm IM=LT+ZONDOT+ZONALS+TIDE+NOISE.}\eqf
 We include in our model the
possibility of varying the length of the time series $T_{\rm obs
}$, the temporal step $\Delta t$ which simulates the orbital arc
length, the amplitude of the noise and of the mismodelling in the
perturbations and the initial phases of the sinusoidal terms in
order to simulate different initial conditions and uncertainties
in the dynamical force models of the orbital processors.
%More
%precisely, the magnitude of the mismodelling in the various
%effects is randomly varied within the currently accepted ranges
%(1-sigma) by using random numbers generated from a normal
%distribution with mean zero, variance one and standard deviations
%one. The same also holds for ZONALS because it is impossible to
%know, a priori, the sign of the slope of the residual even zonal
%trend.
The so-built IM represents the basis of our subsequent analyses.

In order to check the reliability of the adopted procedure we
first try to obtain some of the quantitative features of the real
signal as released by Ciufolini and Pavlis (2004). For example,
they fitted the `raw' curve of Fig. 2 (a) of their paper with a
straight line only obtaining a Root Mean Square (RMS) of the
post-fit residuals of 15 mas. To this aim we perform 5000 runs for
$T_{\rm obs}=11$ years and $x=4$ (EIGEN-GRACE02S) by fitting the
simulated signals with a straight line (LF) $a_0+St$. We calculate
the RMS of the residuals about the straight line fit. The averaged
RMS amounts to 15.6 mas. In Fig. \ref{simulIM}, we plot, among
other things, the simulated time series and the fitted straight
line for one such run representing a generic set of initial
conditions and mismodelling in the force models. It turns out that
the RMS for such IM is 14.8 mas. This shows that our strategy
represents a good starting point.

The recovery of the slope $S$ with LF is reliable because over
5000 runs the averaged error amounts to
\eqi\left\langle\left(\rp{\sigma_{S_{\rm LF}}}{S_{\rm LF
}}\right)100\right\rangle_{\rm 5000\ runs}^{T_{\rm obs}=11\ {\rm
years} }=0.8\%.\eqf

\subsection{Linear and quadratic fits}
In regard to the determined slope of the fitted data, a departure of $\sim
7\%$ from the predicted Lense-Thirring effect, which is present in
IM, occurs for $x=4$ (EIGEN-GRACE02S), i.e. \eqi\left\langle\left|\rp{S_{\rm LF }-S_{\rm LT
}}{S_{\rm LT}}\right|100\right\rangle^{T_{\rm obs}=11\ {\rm years}
}_{\rm 5000\ runs }\sim 7\%.\eqf
The same holds for $x=3.8$ (EIGEN-CG03C). For $x=6.3$ (EIGEN-CG01C) and $x=8.7$ (GGM02S) the departures
amount to $\sim 8\%$ and $\sim 10\%$, respectively. Instead, Ciufolini and Pavlis
(2004) claim that the slope of their linear fit amounts to 47.4
mas yr$^{-1}$, i.e. a $\sim 1\%$ departure from the predictions of
GRT.

%Now we describe how we evaluate the impact of the secular rates of
%the even zonal harmonics on the measurement of the Lense-Thirring
%effect.
In the same set of 5000 runs, we also fit the basic IM, i.e. without any rescaling, with a quadratic
polynomial (QF) $a_0+St+Qt^2$, finding
\eqi\left\langle\left|\rp{S_{\rm QF}-S_{\rm LT }}{S_{\rm
LT}}\right|100\right\rangle^{T_{\rm obs}=11\ {\rm years} }_{\rm
5000\ runs }\sim 14\%.\eqf

\begin{figure}
\begin{center}
\includegraphics[width=13cm,height=11cm]{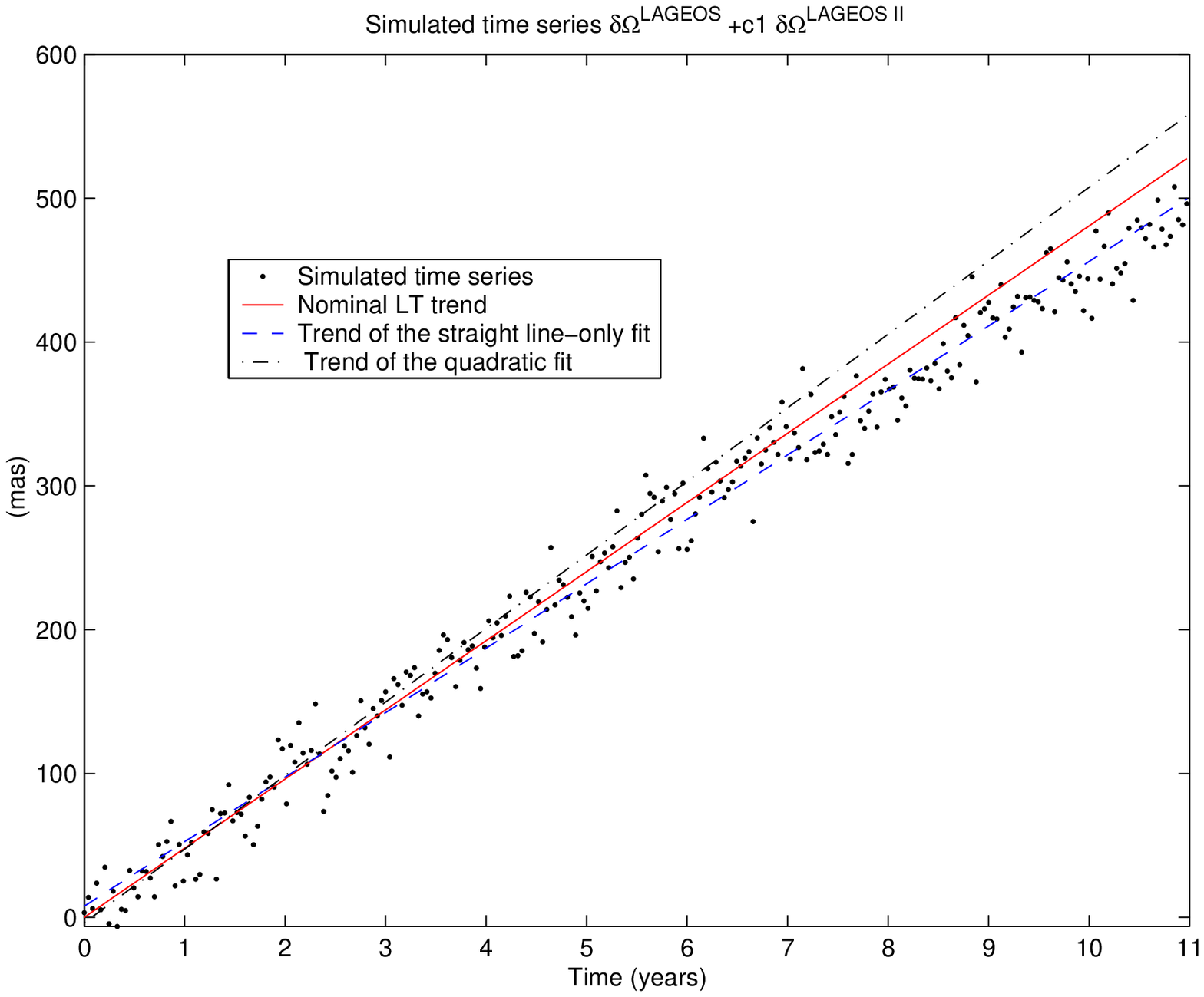}
\end{center}
\caption[Simulated time series , predicted Lense-Thirring trend,
trend of the linear fit and trend of the quadratic fit over 11
years.]{\label{simulIM} Basic, not-rescaled simulated time series
($\boldsymbol{\cdot}$), predicted Lense-Thirring trend
($\overline{\ \ }$), trend of the straight line-only fit ($-\  \
-$) and trend of the quadratic fit ($\cdot\ -\ \cdot$) for $T_{\rm
obs}=11$ years, $\Delta t=15$ days and EIGEN-GRACE02S. The effect of $\dot J_{\ell}$ is
present at 1-sigma level,  according to Table
\ref{jdots}, in the simulated time series. The difference between the slopes of the two fitted
linear trends amount to 13$\%$ of the Lense-Thirring effect. A $7\%$ discrepancy between the straight-line only fit and the predicted LT slope occurs. See Table \ref{conc1}. }
\end{figure}

In order to assess the impact of the secular variations of the
even zonal harmonics, we make the following preliminary remark.
From a simple visual inspection of the plot of the simulated IM in
Fig. \ref{simulIM}, which refers to a IM built with 1-$\sigma_{\dot
J_{\ell}}$, it is not possible to infer anything about the
presence or not of a quadratic term attributable to the secular
variations of the even zonal harmonics. The parabolic signal is, indeed, very smooth and tends to
affect the linear signal, especially as far as the early data are concerned.
%The same would also hold  for 2-sigma and
%3-sigma, as clearly shown in Fig. \ref{jdota} in which we have
%plotted the IM built with $2\sigma_{\dot J_{\ell}}$ and
%$3\sigma_{\dot J_{\ell}}$ for the same set of initial conditions
%and mismodelling.
%In the following, we will work at 1-sigma level.
To be more quantitative, we assess the systematic error due to the
rates of the even zonals by calculating for every run the
difference between the slopes obtained with the previously
described LF and  QF. Note that we do not include the sinusoidal
perturbations in the fits because they are a minor problem. For
example, the $\ell=2, m=0$ 18.6-year tide does not affect
\rfr{iorform} and over 11 years all the most relevant
perturbations describe many full cycles; the longest period is
that of the solar $K_1$ tide perturbation on the node of LAGEOS,
i.e. 2.85 years. Note that the approach of
comparing the slopes of the fits performed with and without the
features of which we want to assess the effect is the same adopted
by Pavlis and Iorio (2002) and Ciufolini and Pavlis (2004) for the
time-dependent harmonic perturbations. However, the problem of the
secular rates of the even zonals was not treated in Ciufolini and
Pavlis (2004), which only dealt with the time-dependent sinusoidal
perturbations finding a 2$\%$ maximum error.

We get \eqi\left\langle\left|\rp{S_{\rm
QF }-S_{\rm LF }}{S_{\rm LT}}\right|100\right\rangle^{T_{\rm
obs}=11\ {\rm years} }_{\rm 5000\ runs }\sim 13\%.\eqf Such results depend on the size of the parabolic signal
introduced in the simulated IM. Indeed, if we conservatively triple the  $\sigma_{\dot J_{\ell}}$ of Table \ref{jdots}
we get a discrepancy of $15\%$. The complete results of our simulations are shown in Tables
\ref{conc1}, \ref{conc3} and \ref{conc5} for  1-$\sigma_{\dot J_{\ell}}$, 3-$\sigma_{\dot J_{\ell}}$ and 5-$\sigma_{\dot J_{\ell}}$, respectively. In Fig.
\ref{simulIM} we plot the linear components of LF and QF for
one of such runs.
%For
%$2\sigma_{\dot J_{\ell}}$ and $3\sigma_{\dot J_{\ell}}$ we obtain
%15$\%$ and 17$\%$, respectively.

\begin{figure}
\begin{center}
\includegraphics[width=13cm,height=11cm]{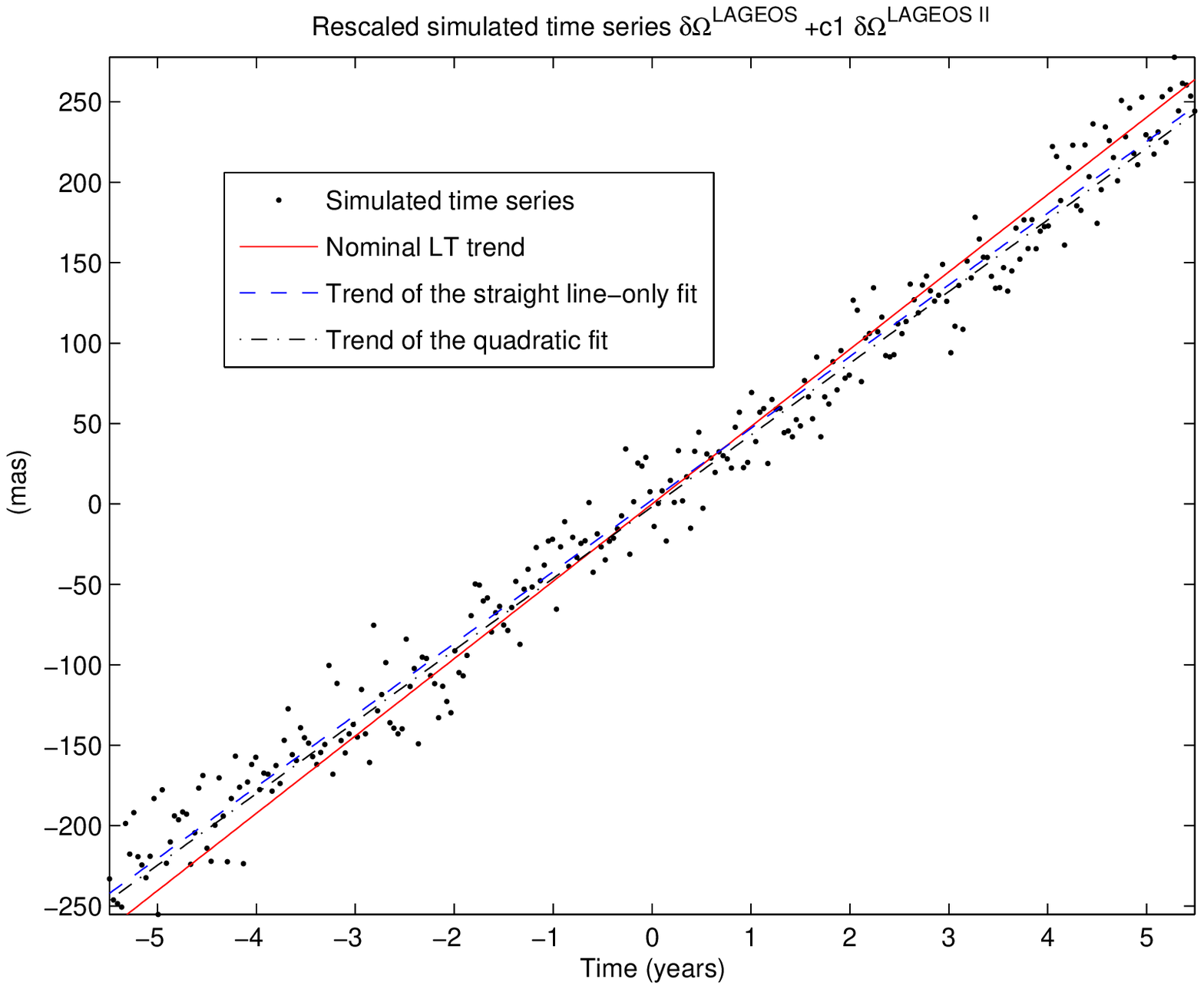}
\end{center}
\caption[Rescaled simulated time series , predicted Lense-Thirring trend,
trend of the linear fit and trend of the quadratic fit over 11
years.]{\label{simulIM2} Rescaled simulated time series
($\boldsymbol{\cdot}$), predicted Lense-Thirring trend
($\overline{\ \ }$), trend of the straight line-only fit ($-\  \
-$) and trend of the quadratic fit ($\cdot\ -\ \cdot$) for $T_{\rm
obs}=11$ years, $\Delta t=15$ days and EIGEN-GRACE02S. The $\dot J_{\ell}$ effect is
present at 1-sigma level,   according to Table
\ref{jdots}, in the simulated time series. The difference between the slopes of the two fitted
linear trends is now negligible. A $7\%$ discrepancy with the predicted LT slope occurs. See Table \ref{Conc}. }
\end{figure}

The same procedure has been repeated by rescaling the data of IM in order to
shift the zero point of the time series in the middle of the data span.
As a result, for $T_{\rm obs}=11$ years the slopes of LF remain unchanged, showing the same $7-10\%$ discrepancy
with respect to the GTR value as before according to the different Earth gravity models used,
while the slopes of QF do change in such a way that they are now identical to those of LF.
In this case the magnitude of $\sigma_{\dot J_{\ell}}$ adopted in the IM does not influence the results.
In Fig.
\ref{simulIM2} we plot the rescaled IM, the Lense-Thirring trend predicted by GTR and the linear
components of LF and QF for
one run.
The results for different time spans and Earth's gravity models are shown in Table \ref{Conc}.

%\subsubsection{The $\dot J_4^{\rm eff}$.}
%%Another way to assess the effect of  $\dot J_{\ell}$ could be, in
%%principle, the following. After fitting IM with QF, we may get a
%%IM$^{'}$ by subtracting the quadratic term $Qt^2$, obtained with
%%QF, from IM. Thus, we may fit the so-obtained `cleaned' time
%%series IM$^{'}$ with a straight line only (LF$^{'}$) and compare
%%the so-obtained slope with that previously obtained with LF.
%%Unfortunately, this procedure is unreliable because it turns out
%Note that the estimated coefficient $Q$ of QF must be considered
%as undetermined in the sense that the errors in the fits are
%always larger than the estimated values: their mean amounts to
%$\sim 190\%$. Also the distribution of the results of the 5000
%runs has a width much larger than its mean value (see Fig.
%\ref{j4doteff}: the percent error calculated with the standard
%deviation of the mean is $\sim 300 \%$). Moreover, such results,
%i.e. the mean estimated value, the mean error and the standard
%deviation of the mean, are not stable and change if one performs
%various numerical experiments for 5000 runs. Incidentally, this
%also rules out the possibility of using the combination of
%\rfr{iorform} for measuring a $\dot J_4^{\rm eff}$, contrary to
%what is claimed by Ciufolini (2004).
%

\section{Discussions and conclusion}
In this paper we addressed the problem of a quantitative
evaluation of the impact of the current lingering uncertainty in the static and varying parts $J_{\ell}$ and
$\dot J_{\ell}$
of the even zonal harmonics of the terrestrial gravitational field on the measurement of the general relativistic Lense-Thirring
effect with the combination of \rfr{iorform} involving the nodes
of LAGEOS and LAGEOS II.

To this aim, we realistically simulated the time series of the LAGEOS satellites data and we performed 5000 runs randomly varying the
noise level, the initial conditions and the mismodelling of the
included dynamical effects within the range of the currently known
uncertainties. In each of such runs we fitted the basic simulated signal
with a straight line and with a quadratic polynomial. The same procedure was also repeated by manipulating
the simulated data rescaling them so that the zero point falls in the middle of the data set.
%and compared
%the slopes of the so-obtained linear trends. The average of such
%differences is assumed to represent a measure of the impact of
%$\dot J_{\ell}$ on the Lense-Thirring effect for a given $T_{\rm
%obs}$.

%Here we summarize the abbreviations used in the text and in Table
%\ref{conc}.
%\begin{itemize}
  %\item []
  %IM. Simulated time series of $\delta\Omega^{\rm LAGEOS}+c_1\delta\Omega^{\rm LAGEOS \ II}$.
  %\item []
  %LF. Linear fitting function $a_0+St$.
  %\item []
  %QF. Quadratic fitting function $a_0+St+Qt^2$.
  %\item []
  %$\Delta_{\rm LF-QF}$. Averaged percent difference between the slopes of LF and
  %QF of the complete IM. It yields the systematic error due to the
  %$\dot J_{\ell}$.
  %\item []
  %$\Delta_{\rm LF-LT}$. Averaged percent difference between the slope of LF of
  %the complete IM and the Lense-Thirring effect as predicted by
  %GRT.
  %%\item []
  %%$\Delta_{\rm QF-LT}$. Averaged percent difference between the slope of QF of
  %%the complete IM and the Lense-Thirring effect as predicted by
  %%GRT.
  %%\item []
  %%$\Delta_{\rm LF-LF^{'}}$. Averaged percent difference between the slope of LF
  %%of the complete IM and the slope of the LF$^{'}$ of the IM to which
  %%the quadratic term obtained in QF has been subtracted.
  %%\item []
  %%$\Delta_{\rm QF-LF^{'}}$. Averaged percent difference between the slope of QF
  %%of the complete IM and the slope of the LF$^{'}$ of the IM to which
  %%the quadratic term obtained in QF has been subtracted.
%\end{itemize}
%
\subsection{The total error budget}
We propose the following error budgets for a data span of 11 years (all the obtained figures are 1-sigma bounds)
\begin{itemize}
\item
From a-priori analytical calculation (EIGEN-GRACE02S)
\eqi\delta\mu_{\rm LT}^{\rm
total}\equiv\sqrt{({\rm grav})^2+
  ({\rm nongrav})^2}\sim\sqrt{(4+12+2)^2+2^2}=18\%.\eqf
  The same also holds for EIGEN-CG03C.
For EIGEN-CG01C we have \eqi\delta\mu_{\rm LT}^{\rm
total}\equiv\sqrt{({\rm grav})^2+
  ({\rm nongrav})^2}\sim\sqrt{(6.3+12+2)^2+4^2}=20\%.\eqf
GGM02S yields \eqi\delta\mu_{\rm LT}^{\rm total}\equiv\sqrt{({\rm
grav})^2+
  ({\rm nongrav})^2}\sim\sqrt{(8.7+12+2)^2+2^2}=23\%.\eqf
 We have included in
our evaluations also the errors due to the tides (2$\%$) and the
non-gravitational perturbations (2$\%$).
\item
From numerical simulations and fits of the basic simulated data (EIGEN-GRACE02S)
\eqi\delta\mu_{\rm LT}^{\rm
total}\equiv\sqrt{({\rm grav})^2+
  ({\rm nongrav})^2}\sim\sqrt{(7+13+2)^2+2^2}=22\%.\eqf
  The same also holds for EIGEN-CG03C.
For EIGEN-CG01C we have \eqi\delta\mu_{\rm LT}^{\rm
total}\equiv\sqrt{({\rm grav})^2+
  ({\rm nongrav})^2}\sim\sqrt{(8+13+2)^2+2^2}=23\%.\eqf
GGM02S yields \eqi\delta\mu_{\rm LT}^{\rm total}\equiv\sqrt{({\rm
grav})^2+
  ({\rm nongrav})^2}\sim\sqrt{(10+13+2)^2+2^2}=25\%.\eqf
\item
From numerical simulations and fits of the rescaled simulated data (EIGEN-GRACE02S)
\eqi\delta\mu_{\rm LT}^{\rm
total}\equiv\sqrt{({\rm grav})^2+
  ({\rm nongrav})^2}\sim\sqrt{(7+2)^2+2^2}=9\%.\eqf
  The same also holds for EIGEN-CG03C.
For EIGEN-CG01C we have \eqi\delta\mu_{\rm LT}^{\rm
total}\equiv\sqrt{({\rm grav})^2+
  ({\rm nongrav})^2}\sim\sqrt{(8+2)^2+2^2}=10\%.\eqf
GGM02S yields \eqi\delta\mu_{\rm LT}^{\rm total}\equiv\sqrt{({\rm
grav})^2+
  ({\rm nongrav})^2}\sim\sqrt{(10+2)^2+2^2}=12\%.\eqf
\end{itemize}
In regard to the evaluation of the error of gravitational origin,
the secular variations of the even zonal harmonics were not solved
for in the latest Earth gravity field models based on CHAMP and
GRACE: they were held fixed to default values obtained from
multi-year data analysis of the data of the constellation of
laser-ranged geodetic satellites. Thus, the recovered values for
$J_{\ell}$ are not independent of $\dot J_{\ell}$.
%In GGM02S,
%$\dot J_2=-2.6\times 10^{-11}$ yr$^{-1}$ only is present, while in
%the GFZ models $\dot J_2=-2.6\times 10^{-11}$ yr$^{-1}$ and $\dot
%J_4=-1.41\times 10^{-11}$ yr$^{-1}$ are included.
Thus, caution advices to linearly sum the biases of the static
part of the geopotential and of their secular variations in
assessing the total systematic error of gravitational origin in
the Lense-Thirring test.

Instead,  Ciufolini and Pavlis (2004) at
the end of the Section Total uncertainty, pag. 960,  and in the
Supplementary Information .doc file add in quadrature the doubled
error due to the static part of the geopotential (i.e. 2$\times
4\%$ value obtained from the sum of the individual error terms),
their optimistic evaluation of the error due to the time dependent
part of the Earth gravity field (2$\%$, not doubled) and the non-gravitational
error (2$\%$, not doubled) getting $\sqrt{8^2+4^2+4^2}\%=10\%$.
% at 2-$\sigma$.
On the other hand, in the Supplementary Information .doc file it
seems that they triple the 3$\%$ error due to the static part of
the even zonal harmonics obtained optimistically with a root-sum-square
calculation and add it in quadrature to the other (not tripled)
errors getting $\sqrt{9^2+2^2+2^2}\%\leq 10\%$.
%at 3-$\sigma$.

These considerations and our results show that the 1-sigma 5$\%$
total error claimed by Ciufolini and Pavlis (2004), who used only
one Earth's gravity model, is optimistic. Moreover, the
present-day uncertainty in the knowledge of the static part of the
geopotential does not yet allow for a fully reliable,
model-independent determination of the gravitomagnetic effect with
the combination of \rfr{iorform}. It should also be considered
that the evaluations presented here might turn out to be
optimistic because they are based on the smoothed values of Table
\ref{jdots}. The impact of possible interannual variations of
$J_4$ and $J_6$, depending also on the particular observational
time span considered, has not been considered here and should
deserve further investigations.

An inspection of Tables \ref{conc1}-\ref{conc5} shows that the secular
variations of the even zonal harmonics may represent a limiting factor
also for longer and shorter time spans.
%This fact contradicts the
%statement by Iorio (2004) according to which an observational time
%span of just a few years would be helpful in reducing the impact
%of $\dot J_{\ell}$.
Unfortunately, the expected improvements in
the knowledge of the static part of the geopotential from the
forthcoming GRACE solutions will not ameliorate the situation with
respect to the problem of the $\dot J_{\ell}$.
\subsection{How to improve the accuracy and the reliability of the Lense-Thirring tests}
Among possible alternatives the most promising one is
  the launch of another satellite of LAGEOS-type, like LARES or OPTIS (L\"{a}mmerzahl et al. 2004). As shown
  in Iorio (2005b), the improvements in our knowledge of the gravitational field from the
  GRACE mission would allow to make the requirements on the orbital geometry of such a new laser target
  much less stringent than
  in the originally proposed LARES mission. In particular, by combining the data of the new spacecraft with those
  of LAGEOS and LAGEOS II, it
  would be possible to reduce its semimajor axis from 12270 km to
  7500-8000 km: this would allow to greatly reduce the costs of the mission.
  Also the inclination could experience large departures from the
  value of LARES ($i=70$ deg) without affecting the outcome of the experiment.
  The accuracy in measuring the Lense-Thirring effect with a three nodes combination including also
  LAGEOS and LAGEOS II
  is $\sim 1\%$ at 1-sigma level.

\newpage
\section*{Acknowledgments}
I thank the anonymous referees due to their efforts which greatly
improved the manuscript.

\newpage
%\section{Tables}
%
{\small\begin{table}[Orbital parameters of the LAGEOS satellites.
]\caption{Orbital parameters of the existing LAGEOS and LAGEOS II
satellites and their Lense-Thirring node precessions in mas
yr$^{-1}$. }\label{para}

\begin{tabular}{lllll}
\noalign{\hrule height 1.5pt}

Satellite & $a$ (km) & $e$ & $i$ (deg) & $\dot\Omega_{\rm LT}$ (mas yr$^{-1}$)  \\

\hline

LAGEOS    &  12270    & 0.0045 &  110 & 31\\
LAGEOS II &  12163    & 0.0135 & 52.64 & 31.5\\
%LARES    & 12270    & 0.04  & 70 & 31\\

\noalign{\hrule height 1.5pt}
\end{tabular}

\end{table}}
{\small\begin{table}\caption[Node classical precession
coefficients ]{Coefficients $\dot\Omega_{.\ell}$ of the node
classical even zonal precessions of LAGEOS and LAGEOS II up to
degree $\ell=20$ in mas yr$^{-1}$. }\label{echecaz}

\begin{tabular}{lll}
\noalign{\hrule height 1.5pt}

$\ell$ & LAGEOS  & LAGEOS II\\

\hline

2 & $4.191586788514\times 10^{11}$ & $-7.669274920758\times 10^{11}$\\
4 & $1.544030247472\times 10^{11}$ & $-5.58637864293\times 10^{10}$\\
6 & $3.25092246054\times 10^{10}$& $4.99185703735\times 10^{10}$\\
8 & $2.1343038821\times 10^{9}$ & $1.10707933989\times 10^{10}$\\
10 &$-1.4885315218\times 10^9$ & $-2.2176133068\times 10^9$\\
12 & $-7.703165634\times 10^{8}$& $-1.1555006405\times 10^9$\\
14 & $-2.097322521\times 10^8$ & $2.5803602\times 10^6$\\
16 & $-3.04891722\times 10^7$ & $8.81906969\times 10^7$\\
18 & $2.7037212\times 10^6$& $1.25437446\times 10^7$\\
20 & $3.3458376\times 10^6$& $-4.8988704\times 10^6$\\
\noalign{\hrule height 1.5pt}
\end{tabular}

\end{table}}

{\small\begin{table}\caption[Systematic gravitational errors
according to three Earth's gravity models. ]{Uncertainty of the
systematic error $\delta\mu_{\rm LT}^{\rm geopot}$ due to the
static part of the even zonal harmonics on the Lense-Thirring
effect for the combination of \rfr{iorform} according to the
variance matrices of the Earth gravity models EIGEN-CG03C,
EIGEN-GRACE02S, EIGEN-CG01C and GGM02S. The linear sum of the
absolute values of the individual terms, according to \rfr{rappos}
with $\sigma_{J_{\ell}}$ instead of $J_{\ell}$, have been used in
order to give realistic 1-sigma upper bounds.}\label{geopo}

\begin{tabular}{lllll}
\noalign{\hrule height 1.5pt}

& EIGEN-CG03C & EIGEN-GRACE02S & EIGEN-CG01C & GGM02S \\

\hline

$\delta\mu_{\rm LT}^{\rm geopot}$ & 3.8$\%$ & 4.1$\%$ & 6.3$\%$ & 8.7$\%$\\

\noalign{\hrule height 1.5pt}
\end{tabular}

\end{table}}
{\small\begin{table}[Secular rates of the even zonal harmonics.
]\caption{The reported values for $\dot J_2,\dot J_4$ and $\dot
J_6$ are the weighted means of the best estimates of Table 1 by
Cox et al. (2003) in units of $10^{-11}$ yr$^{-1}$. The
uncertainties $\sigma_{\dot J_{\ell}}$ are the variances of the
distributions of the formal errors of Table 1 by Cox et al. (2003)
in the same units. }\label{jdots}
\begin{tabular}{llll}
\noalign{\hrule height 1.5pt}

 & $\ell=2$ & $\ell=4$ & $\ell=6$  \\

\hline $\dot J_{\ell}$ & -2.113 & -0.6992 & -0.3594\\
$\sigma_{\dot J_{\ell}}$ & 0.0810 & 0.2029 & 0.1765\\

\noalign{\hrule height 1.5pt}
\end{tabular}

\end{table}}
\begin{table}\caption[Summary of the obtained results]{Results of our tests
for the evaluation of the 1-sigma level impact of $\dot J_{\ell}$
on the measurement of the Lense-Thirring effect with the
combination of \rfr{iorform}. No rescaling has been applied to the simulated time series.
$\Delta_{\rm LF-QF}$ is the averaged
difference between the slopes  of the fitted linear trends in the
linear fit-only (LF) and the quadratic fit (QF). $\Delta_{\rm
LF-LT}$ is the averaged difference between the slopes of the
fitted linear trend in the linear fit-only (LF) and the
Lense-Thirring trend as predicted by GRT. $\Delta_{\rm QF-LT}$ is
the averaged difference between the slopes of the linear part of
the quadratic fit (QF) and the Lense-Thirring trend as predicted
by GRT. A time step
of $\Delta t=15$ days and EIGEN-GRACE02S have been used.}

\label{conc1}

\begin{tabular}{llll}
\noalign{\hrule height 1.5pt}

$T_{\rm obs}$ (yr) & $\Delta_{\rm LF-QF}$ $(\%)$& $\Delta_{\rm LF-LT}$ $(\%)$ & $\Delta_{\rm QF-LT}$ $(\%)$\\
% $\Delta_{\rm QF-LT}$ (\%)\\
\hline
3  &  22     &   11 & 24\\
%    &   24.2\\
6  & 10  & 8 & 14\\
%   & 13.8\\
11 &  13   & 7 & 14\\
%    & 14\\
20 &  12  & 7 & 11\\
%  & 11.3\\
30 &   9  &  9 & 4\\
%   &  4 \\

\noalign{\hrule height 1.5pt}
\end{tabular}

\end{table}
\begin{table}\caption[Summary of the obtained results]{Results of our tests
for the evaluation of the 3-sigma level impact of $\dot J_{\ell}$
on the measurement of the Lense-Thirring effect with the
not-rescaled simulated time series.
A time step
of $\Delta t=15$ days and EIGEN-GRACE02S have been used.}

\label{conc3}

\begin{tabular}{llll}
\noalign{\hrule height 1.5pt}

$T_{\rm obs}$ (yr) & $\Delta_{\rm LF-QF}$ $(\%)$& $\Delta_{\rm LF-LT}$ $(\%)$ & $\Delta_{\rm QF-LT}$ $(\%)$\\
% $\Delta_{\rm QF-LT}$ (\%)\\
\hline
3  &    22   &  10  & 25 \\
%    &   24.2\\
6  & 11  & 9 & 14 \\
%   & 13.8\\
11 &   16  & 10 & 14 \\
%    & 14\\
20 &  18  & 14 & 11\\
%  & 11.3\\
30 &   20  & 20  & 4\\
%   &  4 \\

\noalign{\hrule height 1.5pt}
\end{tabular}

\end{table}
\begin{table}\caption[Summary of the obtained results]{Results of our tests
for the evaluation of the 5-sigma level impact of $\dot J_{\ell}$
on the measurement of the Lense-Thirring effect with the
not-rescaled simulated time series.
A time step
of $\Delta t=15$ days and EIGEN-GRACE02S have been used.}

\label{conc5}

\begin{tabular}{llll}
\noalign{\hrule height 1.5pt}

$T_{\rm obs}$ (yr) & $\Delta_{\rm LF-QF}$ $(\%)$& $\Delta_{\rm LF-LT}$ $(\%)$ & $\Delta_{\rm QF-LT}$ $(\%)$\\
% $\Delta_{\rm QF-LT}$ (\%)\\
\hline
3  &   22   &  11  & 24\\
%    &   24.2\\
6  & 12  & 11 & 14\\
%   & 13.8\\
11 &  18   & 14 & 14\\
%    & 14\\
20 &  25  & 23 & 11\\
%  & 11.3\\
30 &   33  & 33  & 5\\
%   &  4 \\

\noalign{\hrule height 1.5pt}
\end{tabular}

\end{table}
\begin{table}\caption[Summary of the obtained results]{Results of our tests
 with the
rescaled simulated time series.
$\Delta_{\rm LF-QF}$ is now zero and $\Delta_{\rm LF-LT}$ and $\Delta_{\rm QF-LT}$ are equal and denoted as $\Delta$ (in percent). There is no dependence on the size of $\sigma_{\dot J_{\ell}}$. A time step
of $\Delta t=15$ days and different Earth's gravity models have been used.}

\label{Conc}

\begin{tabular}{llll}
\noalign{\hrule height 1.5pt}

$T_{\rm obs}$ (yr) & $\Delta$ (EIGEN-GRACE02S) & $\Delta$ (EIGEN-CG01C) & $\Delta$ (GGM02S) \\
% $\Delta_{\rm QF-LT}$ (\%)\\
\hline
3  &  10 & 12 & 13\\
%    &   24.2\\
6  &  9 &  10 & 11\\
%   & 13.8\\
11 &  7 & 8 & 10\\
%    & 14\\
20 &  4 & 5 & 7\\
%  & 11.3\\
30 &   3 &  5& 7\\
%   &  4 \\

\noalign{\hrule height 1.5pt}
\end{tabular}

\end{table}

\newpage

%\section{Figures}

%\begin{figure}
%\begin{center}
%\includegraphics[width=13cm,height=11cm]{jdots.eps}
%\end{center}
%\caption{\label{jdota} Simulated IM by assuming 1,2,3-sigma
%mismodelling in the secular rates of the even zonal harmonics.
%$T_{\rm obs}=11$ years, $\Delta t=15$ days, $\sigma_{J_{\ell}}$
%from EIGEN-GRACE02S, $\sigma_{\dot J_{\ell}}$ from Table
%\ref{jdots}. }
%\end{figure}

%\begin{figure}
%\begin{center}
%\includegraphics[width=13cm,height=11cm]{j4doteff.eps}
%\end{center}
%\caption{\label{j4doteff} Distribution of the recovered values of
%$\dot J_4^{\rm eff}$ from the coefficients $Q$ of the quadratic
%fits over a set of 5000 runs. The standard deviation of the mean
%is 300 $\%$ of the mean value.}
%\end{figure}

\end{document}